\documentstyle[11pt,epsfig]{article}

\title{Confidence Intervals for Poisson Distribution Parameter}

\author{\large 
S.I.~Bityukov$^1$, N.V.~Krasnikov$^2$, V.A. Taperechkina$^3$ \\[3mm]
\em Institute for High Energy Physics, Protvino,  \\
\em Russia}
\date{}

\begin{document}
\maketitle

%\author{Sergei Bityukov}\address{Institute for High Energy Physics, 
%Protvino, 142284, Russia, bityukov@mx.ihep.su} 
%\author{Nikolai Krasnikov}\address{Institute for Nuclear Research RAS,
% Moscow, Russia, krasniko@ms2.inr.ac.ru}
%\author{Vera Taperechkina}\address{Moscow State Academy of Instrument 
%Engineering and Computer Science, Serpukhov, Russia}
%\date{5 August, 2001}

%\begin{abstract}
Results of numerical procedure of constructing confidence intervals
for parameter of the Poisson distribution of signal events in the 
presence of background events with known value of parameter of Poisson 
distribution are presented. It is shown that the used procedure has both 
the Bayesian and frequentist interpretations. Also the possibility to 
construct a continuous analogue of the Poisson distribution to search 
the bounds of confidence intervals for the parameter of the Poisson 
distribution is discussed. 
%\end{abstract}

%\keywords{Gamma distribution, Poisson distribution, confidence intervals}

\vspace{1cm}
\bigskip

\noindent
\rule{3cm}{0.5pt}\\
$^1$~~bityukov@mx.ihep.su,Serguei.Bitioukov@cern.ch \\
$^2$~~Institute for Nuclear Research RAS,  Moscow.\\
$^3$~~Moscow State Academy of Instrument 
Engineering and Computer Science, Serpukhov, Russia.

\section{Introduction}

In paper~\cite{1} the unified approach to the construction
of confidence intervals and confidence limits
for a signal in the background presence, 
in particular, for Poisson distributions, is proposed. The method is widely
used for the presentation of physical results~\cite{2}
though a number of investigators criticize this approach~\cite{3}.

Here we use a simple method of constructing 
confidence intervals for the Poisson distribution parameter for a signal 
in the presence of background which has the Poisson distribution with 
the known value of parameter to compare with a conventional 
procedure~\footnote{The early version of the study can be
found in S.I. Bityukov, N.V. Krasnikov, arXiv:physics/0009064.}~. 
The method is based on the statement that the probability of true value  
of the Poisson distribution parameter to be a specified value
(in the case of the observed number of events $\hat x$) distributes 
in accordance with a Gamma distribution. 
It is shown that this statement
has both Bayesian and frequentist interpretations.
The experimental results often give non-integer values for a number of 
observed events $\hat x$ 
(for example, after the background subtraction~\cite{4})
when the Poisson distribution occurs. That is why there is a necessity 
to have a procedure for constructing the confidence intervals 
in this case. The paper offers a generalization  of Poisson 
distribution for a continuous case~\footnote{The early version of the study 
can be found in S.I. Bityukov, N.V. Krasnikov, V.A. Taperechkina,
arXiv:physics/0008082; also Preprint IHEP 2000-61, Protvino, 2000.}~. 
The generalization given here allows one
to construct confidence intervals and confidence limits for the Poisson 
distribution parameter (for integer and real values of a number of 
observed events) using conventional methods. 

In Sect.~\ref{sect:relation} 
the interrelation between the frequentist and Bayesian definitions
of the confidence interval is shown. The method of constructing confidence 
intervals for the Poisson distribution parameter for a signal in the presence 
of background which has the Poisson distribution with the known value of 
parameter is described in Sect.~\ref{sect:method}. 
The results of confidence intervals construction and their
comparison with the results of the unified approach are also given 
in Sect.~\ref{sect:method}. In Sect.~\ref{sect:general} 
the generalization of Poisson distribution for the
continuous case is introduced. The examples of confidence intervals
construction for the parameter of the Poisson distribution analogue and
for the Poisson distribution parameter using the Gamma distribution
are considered in Sect.~\ref{sect:central} and in Sect.~\ref{sect:confidence}. 
The main results of the paper are summarized in the Conclusion.

\section{The interrelation between frequentist and Bayesian definitions
of confidence interval}
\label{sect:relation}

Let us have a random value $\xi$, taking values from the set of
numbers $x \in X$. Consider the two-dimensional function

$$
%\begin{equation}
f(x,\lambda) =\displaystyle \frac{\lambda^x}{x!} e^{-\lambda}, \eqno (2.1)
%\end{equation}
$$

\noindent
where $x \ge 0$ and $\lambda > 0$.

Assume, that the set $X$ includes only the whole numbers, then for each
value of $\lambda$ a discrete function $f(x,\lambda)$ 
describes the distribution of probabilities for the Poisson distribution
with the parameter $\lambda$ and a random variable $x$,
i.e. $\xi \sim Pois(\lambda)$.

\begin{figure}[ht]
\epsfig{file=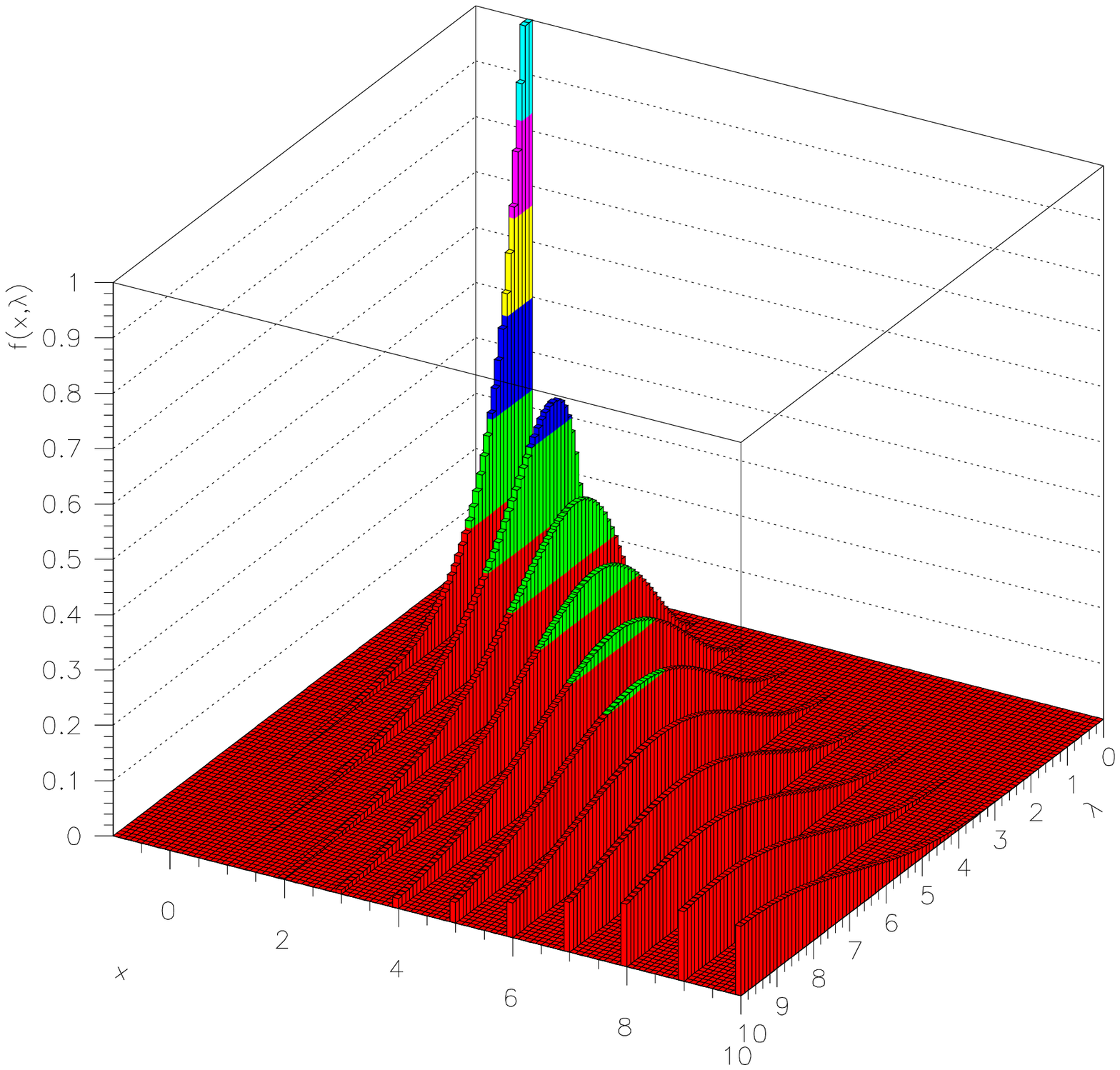,width=15cm}
%  \centering \includegraphics[height=0.4\hsize]{surf2i.eps}
\caption{\small The behaviour of the probability density of the true
     value of parameter $\lambda$ for the Poisson distribution in case 
     of $x$ observed events versus $\lambda$ and $x$. Here 
     $f(x,\lambda) =\displaystyle \frac{\lambda^x}{x!} e^{-\lambda}$ 
     is both the Poisson distribution with the 
     parameter $\lambda$ along the axis $x$
     and the Gamma distribution with a shape parameter $x+1$ and a scale 
     parameter 1 along the axis $\lambda$.}
\label{fig:1}
\end{figure}

Let us write down the density of Gamma distribution $\Gamma_{a,x + 1}$ as

$$
%\begin{equation}
f(x,a,\lambda) = \displaystyle 
\frac{a^{x+1}}{\Gamma(x+1)} e^{-a\lambda} \lambda^{x},  \eqno (2.2)
%\end{equation}
$$

\noindent
where $a$ is a scale parameter, $x + 1 > 0$ is a shape 
parameter, $\lambda > 0$ is a random variable, and $\Gamma(x+1)$ is
a Gamma function. 
Since the $x$ is integer, then $x! = \Gamma(x+1)$. 
Note that this notation is also used in the case of real $x$.
Let us set $a = 1$, then for each $x$ a continuous function

$$
%\begin{equation}
f(x,\lambda) = \displaystyle \frac{\lambda^x}{x!} e^{-\lambda},~ 
\lambda > 0,~x > -1  \eqno (2.3)
%\end{equation}
$$

\noindent
is the density of Gamma distribution $\Gamma_{1, x + 1}$ with the 
scale parameter $a = 1$ 
(see Fig.~\ref{fig:1}). The mean, mode, and variance of this distribution
are given by $x+1,~x$, and $x+1$, respectively.

%%%%%%%%%%%%%%%%%%%%%%%%%%fig.1

Assume that in the experiment with a fixed integral luminosity
(i.e. the process under study is considered as a homogeneous process
for a given time) the $\hat x$ events of a Poisson process
are observed. It means that we have an experimental estimation
$\hat \lambda(\hat x)$ of the parameter $\lambda$ of the Poisson distribution.
We have to construct a confidence interval
$(\hat \lambda_1(\hat x), \hat \lambda_2(\hat x))$, covering
the true value of the parameter $\lambda$ of the distribution under
study with a confidence level $1 - \alpha$, where $\alpha$ is a
significance level. It is known from the theory of statistics~\cite{5},
that the mean value of a sample of data is an unbiased estimation
of the mean of distribution under study. In our case the sample consists
of one~\footnote{The Poisson distributed random values 
have a property: if $\xi \sim Pois(\lambda_1)$ and
$\eta \sim Pois(\lambda_2)$ then 
$\xi + \eta \sim Pois(\lambda_1+\lambda_2)$. It means that if
we have two measurements $\hat x_1$ and $\hat x_2$
of the same random value $\xi \sim Pois(\lambda)$, we can consider
these measurements as one measurement $\hat x_1 + \hat x_2$ of the 
random value $2 \cdot \xi \sim Pois(2 \cdot \lambda)$.}~observation 
$\hat x$.

For the discrete Poisson distribution
the mean coincides with the estimation of parameter value, 
i.e. $\hat \lambda = \hat x$ in our case.

Let us  consider the formula 

$$
%\begin{equation}
P(\lambda|\hat x) = P(\hat x|\lambda) = 
\displaystyle \frac{\lambda^{\hat x}}{\hat x!} e^{-\lambda}. \eqno (2.4)
%\end{equation}
$$

This formula (2.4) results from the Bayesian formula~\cite{6}

$$
%\begin{equation}
P(\lambda|\hat x) P(\hat x) = P(\hat x|\lambda) P(\lambda) \eqno (2.5)
%\end{equation}
$$

\noindent
in the assumption that all the possible values of parameter $\lambda$
have equal probability, i.e. $P(\lambda) = const$.  
In this assumption the probability that unknown parameter $\lambda$
obeys the inequalities $\lambda_1~\le~\lambda~\le~\lambda_2$ 
is given by the evident Bayesian formula 

$$
%\begin{equation}
P(\lambda_1 \le \lambda \le \lambda_2|\hat x) = 
P(\lambda_1 \le \lambda|\hat x) - P(\lambda_2 \le \lambda|\hat x) = 
\int_{\lambda_1}^{\lambda_2}{P(\lambda|\hat x)d\lambda},  \eqno (2.6)
%\end{equation}
$$

\noindent
where $P(\lambda|\hat x)$ is determined by formula (2.4).

Formula (2.6) has also a well defined frequentist meaning.
Using the identity~\footnote{Notice that this identity takes place
for any $\lambda_1 \ge 0$ and $\lambda_2 \ge 0$ (in particular, if
$\lambda_2 \le \lambda_1$).} 

$$
%\begin{equation}
\sum_{i = \hat x + 1}^{\infty}{\frac{\lambda_1^ie^{-\lambda_1}}{i!}} + 
\int_{\lambda_1}^{\lambda_2}
{\frac{\lambda^{\hat x}e^{-\lambda}}{\hat x!}d\lambda} 
+ \sum_{i = 0}^{\hat x}{\frac{\lambda_2^ie^{-\lambda_2}}{i!}} = 1  \eqno (2.7)
%\end{equation}
$$

\noindent
one can rewrite formula (2.6) as

$$
%\begin{equation}
P(\lambda_1~\le~\lambda~\le~\lambda_2|\hat x) = 
1 - P(n \le \hat x|\lambda_2) - P(n > \hat x|\lambda_1) = 
P(n \le \hat x|\lambda_1) - P(n \le \hat x|\lambda_2),  \eqno (2.8)
%\end{equation}
$$

\noindent
where $P(n \le \hat x|\lambda) = \displaystyle
\sum_{n = 0}^{\hat x}{\frac{\lambda^ne^{-\lambda}}{n!}}$ and 
$P(n > \hat x|\lambda) = \displaystyle
\sum_{n = \hat x + 1}^{\infty}{\frac{\lambda^ne^{-\lambda}}{n!}}$.

\begin{figure}
\epsfig{file=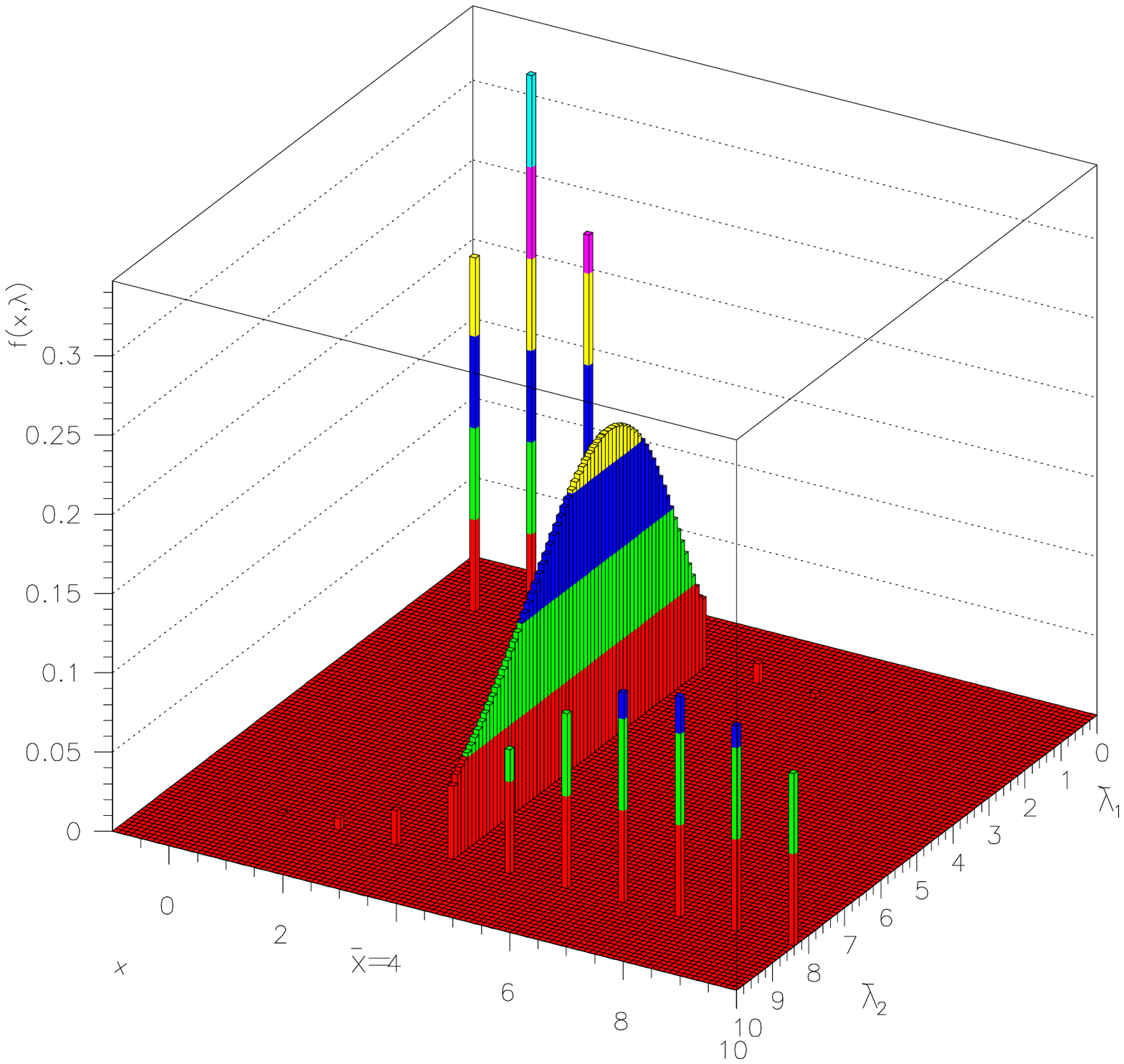,width=15cm}
%  \centering \includegraphics[height=0.4\hsize]{newc.eps}
%\centerline{\psfig{figure=newc.eps,height=6.2cm}}
\caption{\small The Poisson distributions $f(x,\lambda)$ 
for $\lambda$'s determined by the confidence limits $\hat \lambda_1 = 1.51$ 
and $\hat \lambda_2 = 8.36$ in case of the observed number of events 
$\hat x = 4$ are shown. The probability density of Gamma distribution 
with a scale parameter $a=1$ and a shape parameter  
$x+1=\hat x+1=5$ is shown 
within this confidence interval.}
\label{fig:2}
\end{figure}

The right hand side of formula (2.8) has a well defined frequentist
meaning and it is the definition of the
confidence interval in the frequentist approach.
Note, that this definition of the confidence interval for the Poisson 
distribution parameter is self-consistent 
both for the case $\lambda_1 = \lambda_2$
and for the case $\hat x = 0$. 
As an example of the shortest 90\% CL confidence interval of such type
in case of the observed number of events $\hat x=4$ is shown 
in Fig.~\ref{fig:2}.

%%%%%%%%%%%%%%%%%%%%%%%%%%fig.2
 
In ref.~\cite{7}~(pp.406-407) the interrelation between 
the frequentist and Bayesian definitions of confidence interval is shown, 
nevertheless, the author criticizes the 
Bayesian approach of the confidence interval determination. 

As it is seen from the identity~(2.7)
the probability of true value of parameter of Poisson distribution
to be equal to the value of $\lambda$ in the case of one
measurement $\hat x$ has probability density of Gamma distribution 
$\Gamma_{1,1+\hat x}$. 

Correspondingly, in the case of $m$ 
measurements $\hat x_1, \hat x_2, \dots, \hat x_m$ of the 
random values $\xi_1, \xi_2, ..., \xi_m$, where $\xi_i \sim Pois(\lambda)$
for $i = 1, 2, \dots, m$,
the probability of true value of parameter of Poisson distribution
to be equal to the value of $\lambda$ has probability density of 
Gamma distribution $\Gamma_{m,1+\sum_{i = 1}^{m}{\hat x_i}}$. 

\section{The method of confidence intervals construction}
\label{sect:method}

Let us consider the Poisson distribution with two components:
Signal component with a parameter $\lambda_s$ and background component
with a parameter $\lambda_b$, where $\lambda_b$ is known.
To construct confidence intervals for the parameter $\lambda_s$ of a signal 
in the case of observed value $\hat x$, we must find the distribution 
$P(\lambda_s|\hat x)$. 

Firstly let us consider the simplest case $\hat x = \hat s + \hat b = 1$.
Here $\hat s$ is the number of signal events and $\hat b$ is the number of
background events among the observed $\hat x$ events.

The $\hat b$ can be equal to 0 and 1.
We know that the $\hat b$ is equal to 0 with probability 

$$
%\begin{equation}
p_0 = P(\hat b = 0) = 
\displaystyle \frac{\lambda_b^0}{0!} e^{-\lambda_b} = e^{-\lambda_b} \eqno (3.1)
%\end{equation}
$$

and the $\hat b$ is equal to 1 with probability 

$$
%\begin{equation}
p_1 = P(\hat b = 1) = 
\displaystyle \frac{\lambda_b^1}{1!} e^{-\lambda_b} = 
\lambda_b e^{-\lambda_b}.  \eqno (3.2)
%\end{equation}
$$

Correspondingly, 
$P(\hat b = 0|\hat x = 1) = P(\hat s = 1|\hat x = 1) =
\displaystyle \frac{p_0}{p_0 + p_1}$ and
$P(\hat b = 1|\hat x = 1) = P(\hat s = 0|\hat x = 1) =
\displaystyle \frac{p_1}{p_0 + p_1}$.

It means that the distribution of $P(\lambda_s|\hat x = 1)$ 
is equal to the sum of distributions 

$$
%\begin{equation}
P(\hat s = 1|\hat x = 1) \Gamma_{1,2} + 
P(\hat s = 0|\hat x = 1) \Gamma_{1,1}
= \displaystyle \frac{p_0}{p_0 + p_1} \Gamma_{1,2} +
\displaystyle \frac{p_1}{p_0 + p_1} \Gamma_{1,1}, \eqno (3.3)
%\end{equation}
$$

\noindent
where $\Gamma_{1,1}$ is the Gamma distribution with the probability
density $P(\lambda_s|\hat s = 0) = \displaystyle e^{-\lambda_s}$ 
and $\Gamma_{1,2}$ is the Gamma distribution with the probability
density $P(\lambda_s|\hat s = 1) = \displaystyle \lambda_s e^{-\lambda_s}$.
As a result, we have 

$$
%\begin{equation}
P(\lambda_s|\hat x = 1) = 
\displaystyle \frac{\lambda_s + \lambda_b}{1 + \lambda_b} 
\displaystyle e^{-\lambda_s}.  \eqno (3.4)
%\end{equation}
$$

Using formula (3.4) for $P(\lambda_s|\hat x = 1)$ and formula (2.8),
we construct the shortest confidence interval
of any confidence level in a trivial way.

In this manner we can construct the distribution of $P(\lambda_s|\hat x)$
for any values of $\hat x$ and $\lambda_b$. As a result, we have obtained 
the known formula~\cite{8,9}

$$
%\begin{equation}
P(\lambda_s|\hat x) = \displaystyle
\frac{(\lambda_s + \lambda_b)^{\hat x} }
{\hat x! \displaystyle \sum_{i=0}^{\hat x}{\lambda_b^i \over {i!}}}
\displaystyle e^{-\lambda_s}. \eqno (3.5)
%\end{equation}
$$

The numerical results for the
confidence intervals and the results of paper~\cite{1}
are compared in Table~\ref{table:1} and Table~\ref{table:2}.

\setlength{\tabcolsep}{4pt}
\begin{table}
\begin{center}
\caption{90\% C.L. intervals for the Poisson signal mean $\lambda_s$, 
for total events observed $\hat x$, for known mean background $\lambda_b$
ranging from 0 to 2. Comparison between the results of ref.[1] and the 
results from the present paper.}
\label{table:1}
\begin{tabular}{r|cccccccccc} \hline
~$\hat x\backslash\lambda_b$~  & 
    0.0 ref.[1] &  0.0        & 1.0 ref.[1] &    1.0      & 2.0 ref.[1] & 
          2.0        \\ 
\hline
0 &  0.00, 2.44 &  0.00, 2.30 &  0.00, 1.61 &  0.00, 2.30 &  0.00, 1.26 &  
       0.00, 2.30  \\
1 &  0.11, 4.36 &  0.09, 3.93 &  0.00, 3.36 &  0.00, 3.27 &  0.00, 2.53 &  
       0.00, 3.00  \\
2 &  0.53, 5.91 &  0.44, 5.48 &  0.00, 4.91 &  0.00, 4.44 &  0.00, 3.91 &  
       0.00, 3.88  \\
3 &  1.10, 7.42 &  0.93, 6.94 &  0.10, 6.42 &  0.00, 5.71 &  0.00, 5.42 & 
       0.00, 4.93  \\
4 &  1.47, 8.60 &  1.51, 8.36 &  0.74, 7.60 &  0.51, 7.29 &  0.00, 6.60 &  
       0.00, 6.09  \\
5 &  1.84, 9.99 &  2.12, 9.71 &  1.25, 8.99 &  1.15, 8.73 &  0.43, 7.99 &  
       0.20, 7.47  \\
6 &  2.21,11.47 &  2.78,11.05 &  1.61,10.47 &  1.79,10.07 &  1.08, 9.47 &  
       0.83, 9.01  \\
7 &  3.56,12.53 &  3.47,12.38 &  2.56,11.53 &  2.47,11.38 &  1.59,10.53 &  
       1.49,10.37  \\
8 &  3.96,13.99 &  4.16,13.65 &  2.96,12.99 &  3.18,12.68 &  2.14,11.99 &  
       2.20,11.69  \\
9 &  4.36,15.30 &  4.91,14.95 &  3.36,14.30 &  3.91,13.96 &  2.53,13.30 &  
       2.90,12.94  \\
10 & 5.50,16.50 &  5.64,16.21 &  4.50,15.50 &  4.66,15.22 &  3.50,14.50 &  
       3.66,14.22  \\
20 & 13.55,28.52& 13.50,28.33 & 12.55,27.52 & 12.53,27.34 & 11.55,26.52 & 
       11.53,26.34 \\
\end{tabular}
\end{center}
\end{table}

\begin{table}
\begin{center}
\caption{90\% C.L. intervals for the Poisson signal mean $\lambda_s$, 
for total events observed $\hat x$, for known mean background $\lambda_b$
ranging from 6 to 15. Comparison between the results of ref.[1] and 
the results from the present paper.}
\label{table:2}
\begin{tabular}{r|cccccccccc} \hline
~$\hat x\backslash\lambda_b$~  & 
    6.0 ref.[1] &    6.0      &   
            12.0 ref.[1]&   12.0      & 15.0 ref.[1]&   15.0      \\ 
\hline
0 & 0.00, 0.97  & 0.00, 2.30  & 
       0.00, 0.92  & 0.00, 2.30  & 0.00, 0.92  &  0.00, 2.30  \\
1 & 0.00, 1.14  & 0.00, 2.63  &  
       0.00, 1.00  & 0.00, 2.48  & 0.00, 0.98  &  0.00, 2.45  \\
2 &  0.00, 1.57 & 0.00, 3.01  & 
       0.00, 1.09  & 0.00, 2.68  & 0.00, 1.05  &  0.00, 2.61  \\
3 &  0.00, 2.14 &  0.00, 3.48 & 
        0.00, 1.21  & 0.00, 2.91  & 0.00, 1.14  &  0.00, 2.78  \\
4 &  0.00, 2.83 & 0.00, 4.04  & 
        0.00, 1.37  & 0.00, 3.16  & 0.00, 1.24  &  0.00, 2.98  \\
5 &  0.00, 4.07 &  0.00, 4.71 & 
         0.00, 1.58 &  0.00, 3.46 & 0.00, 1.32  &  0.00, 3.20 \\
6 &  0.00, 5.47 &  0.00, 5.49 & 
         0.00, 1.86 &  0.00, 3.80 &  0.00, 1.47 & 0.00, 3.46  \\
7 &  0.00, 6.53 &  0.00, 6.38 & 
         0.00, 2.23 &  0.00, 4.19 &  0.00, 1.69 &  0.00, 3.74 \\
8 &  0.00, 7.99 &  0.00, 7.35 & 
         0.00, 2.83 &  0.00, 4.64 &  0.00, 1.95 &  0.00, 4.06 \\
9 &  0.00, 9.30 &  0.00, 8.41 & 
        0.00, 3.93 &  0.00, 5.15 &  0.00, 2.45 &  0.00, 4.42 \\
10 &  0.22,10.50 &  0.02, 9.53 & 
         0.00, 4.71 &  0.00, 5.73 &  0.00, 3.00 &  0.00, 4.83 \\
20 &  7.55,22.52 &  7.53,22.34 & 
         2.23,16.52 &  1.70,16.08 &  0.00,13.52 &  0.00,12.31 \\
\end{tabular}
\end{center} 
\end{table}

\setlength{\tabcolsep}{1.4pt}

It should be noted that 
in our approach the dependence of the 
confidence intervals width for the parameter $\lambda_s$ 
on the value of $\lambda_b$ in the case $\hat x = 0$ is absent. 
For $\hat x = 0$ the method proposed in ref.~\cite{10} also gives 
a 90\% upper limit independent of $\lambda_b$. This dependence
is also absent in the Bayesian approach~\cite{8,11}.

\section{The Generalization of Discrete Poisson Distribution for
the Continuous Case}
\label{sect:general}

\begin{figure}
\epsfig{file=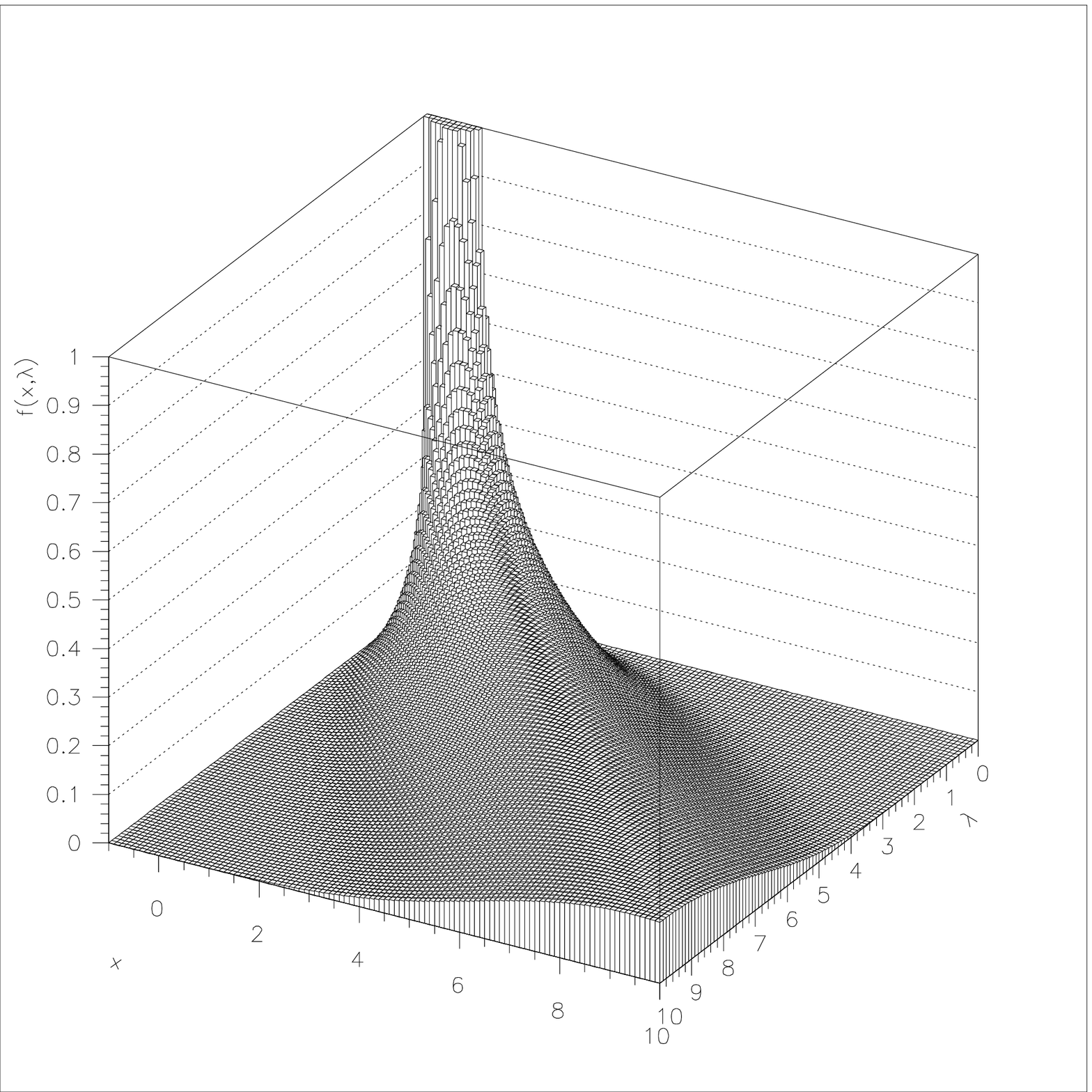,width=15cm}
%  \centering \includegraphics[height=0.4\hsize]{surf2r.eps}
%\centerline{\psfig{figure=surf2r.eps,height=6.2cm}}
\caption{\small The behaviour of the function $f(x,\lambda)$ versus
$\lambda$ and $x$.}
\label{fig:3}
\end{figure}

\begin{figure}[H]
\epsfig{file=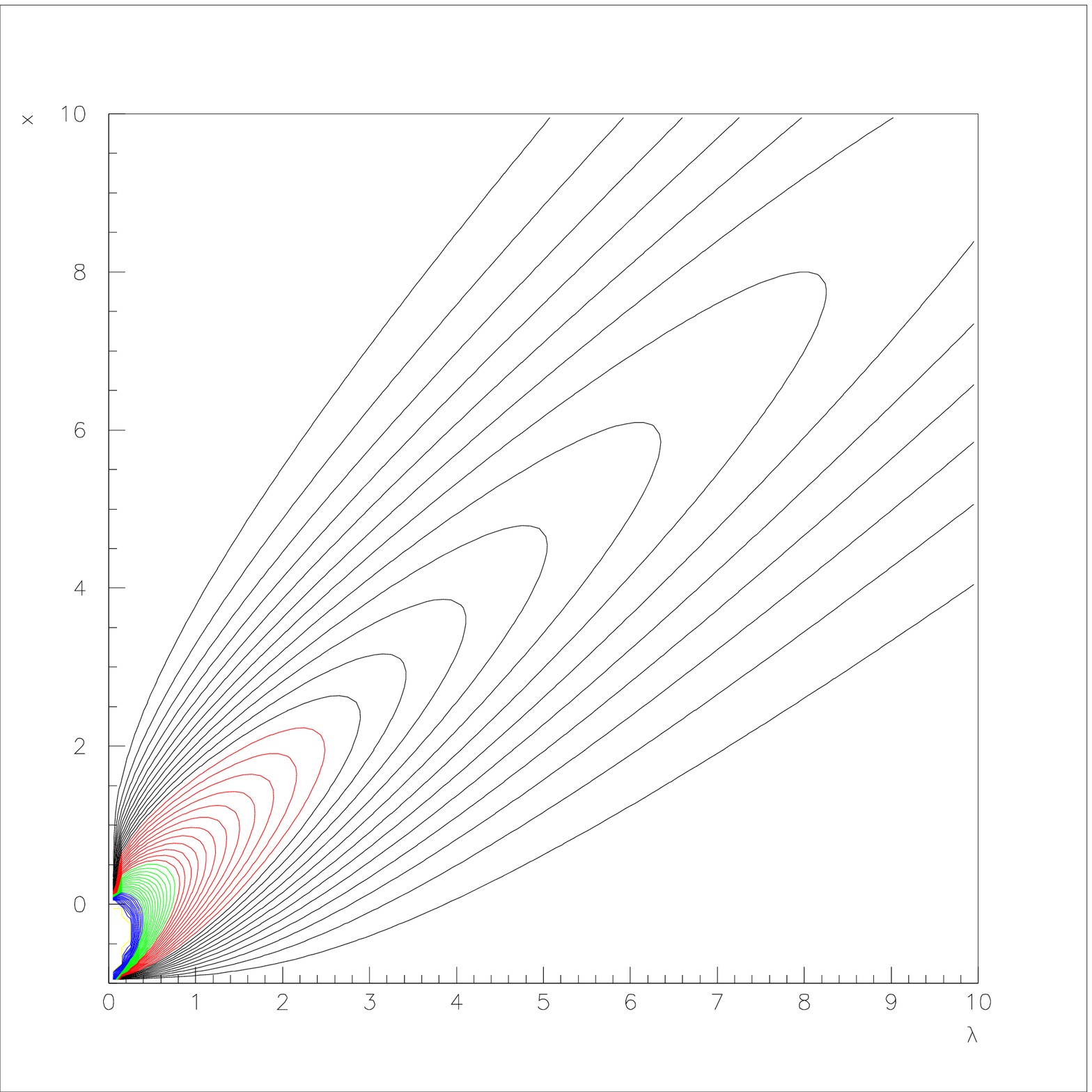,width=15cm}
%  \centering \includegraphics[height=0.4\hsize]{surf3.eps}
%\centerline{\psfig{figure=surf3.eps,height=6.2cm}}
\caption{\small Two-dimensional representation 
of the function $f(x,\lambda)$ versus
$\lambda$ and $x$ for the values $f(x,\lambda) < 1$.}
\label{fig:4}
\end{figure}

Let us consider the case when $x \in X$ are the real values and denote 
$x! = \Gamma(x+1)$, then we can consider the function

$$
%\begin{equation}
f(x,\lambda) =\displaystyle \frac{\lambda^x}{x!} e^{-\lambda}  \eqno (4.1)
%\end{equation}
$$

\noindent
as a continuous two-dimensional function. 
Fig.~\ref{fig:3} shows the surface described by 
this function. Smooth behaviour of this function along
$x$ and $\lambda$ (see Fig.~\ref{fig:4}) allows one 
to assume that there is such a function
$\l(\lambda) > -1$, that 

$$
%\begin{equation}
\displaystyle \int_{l(\lambda)}^{\infty}{f(x,\lambda)dx} = 1  \eqno (4.2)
%\end{equation}
$$

\noindent
for the given value of $\lambda$. It means that in this way we introduce 
a continued 
analogue of Poisson distribution with the probability density
$f(x,\lambda) = \displaystyle \frac{\lambda^x}{x!} e^{-\lambda}$ over the
area of the function definition, i.e. for $x \ge l(\lambda)$ 
and $\lambda > 0$.

%%%%%%%%%%%%%%%%%%%%%%%%%%fig.3

%%%%%%%%%%%%%%%%%%%%%%%%%%fig.4
 
The values of the function $f(x,\lambda)$ for integer $x$ 
coincide with corresponding magnitudes in the probabilities distribution
of discrete Poisson distribution.
Dependences of the values of function $\l(\lambda)$, the means and the
variances for the suggested distribution on $\lambda$ have been calculated
by using the programme DGQUAD from the library  
CERNLIB~\cite{12} and the results are presented in Table~\ref{table:3}.
This Table shows that the series of properties of Poisson distribution
$(E\xi = \lambda, D\xi = \lambda)$ take place only when the value of 
the parameter $\lambda > 3$.  

\setlength{\tabcolsep}{4pt}
\begin{table}
\begin{center}
\caption{The function $l(\lambda)$, mean and variance versus $\lambda$.}
\label{table:3}
\begin{tabular}{|c|c|c|c|}
\hline
$\lambda$ & $l(\lambda)$ & mean $(E\xi)$ & variance $(D\xi)$ \\ 
\hline
    0.001 &  -0.297 &     -0.138  &  0.024  \\
    0.002 &  -0.314 &     -0.137  &  0.029  \\
    0.005 &  -0.340 &     -0.130  &  0.040  \\
    0.010 &  -0.363 &     -0.120  &  0.052  \\
    0.020 &  -0.388 &     -0.100  &  0.071   \\
    0.050 &  -0.427 &     -0.051  &  0.113    \\
    0.100 &  -0.461 &      0.018  &  0.170    \\
    0.200 &  -0.498 &      0.142  &  0.272    \\
    0.300 &  -0.522 &      0.256  &  0.369    \\
    0.400 &  -0.539 &      0.365  &  0.464   \\
    0.500 &  -0.553 &      0.472  &  0.559    \\
    0.600 &  -0.564 &      0.577  &  0.653    \\
    0.700 &  -0.574 &      0.681  &  0.748    \\
    0.800 &  -0.582 &      0.785  &  0.844    \\
    0.900 &  -0.590 &      0.887  &  0.939    \\
     1.00 &  -0.597 &      0.989  &  1.035    \\
     1.50 &  -0.622 &      1.495  &  1.521    \\
     2.00 &  -0.639 &      1.998  &  2.012    \\
     2.50 &  -0.650 &      2.499  &  2.506    \\
     3.00 &  -0.656 &      3.000  &  3.003    \\
     3.50 &  -0.656 &      3.500  &  3.501   \\
     4.00 &  -0.647 &      4.000  &  3.999    \\
     4.50 &  -0.628 &      4.500  &  4.498    \\
     5.00 &  -0.593 &      5.000  &  4.997    \\
     5.50 &  -0.539 &      5.500  &  5.497    \\
     6.00 &  -0.466 &      6.000  &  5.996    \\
     6.50 &  -0.373 &      6.500  &  6.495    \\
     7.00 &  -0.262 &      7.000  &  6.995    \\
     7.50 &  -0.135 &      7.500  &  7.494    \\
     8.00 &   0.000 &      8.000  &  7.993    \\
     8.50 &   0.000 &      8.500  &  8.496    \\
     9.00 &   0.000 &      9.000  &  8.997    \\
     9.50 &   0.000 &      9.500  &  9.498    \\
     10.0 &   0.000 &      10.00  &  9.999    \\
\hline
\end{tabular}
\end{center}
\end{table}

\setlength{\tabcolsep}{1.4pt}

It is appropriate at this point to say that

$$
%\begin{equation}
\displaystyle \int_0^{\infty}{f(x,\lambda)dx} = 
\displaystyle \int_0^{\infty}{\frac{\lambda^xe^{-\lambda}}{\Gamma(x+1)}dx} =
e^{-\lambda}\nu(\lambda). \eqno (4.3)
%\end{equation}
$$

\noindent
The function 

$$
%\begin{equation}
\nu(\lambda) = 
\displaystyle \int_0^{\infty}{\frac{\lambda^x}{\Gamma(x+1)}dx}  \eqno (4.4)
%\end{equation}
$$

\noindent
is well known
and, according to ref.~\cite{13},

$$
%\begin{equation}
\nu(\lambda) = 
\displaystyle \sum_{n=-N}^{\infty}{\frac{\lambda^n}{\Gamma(n+1)}}
+ O(|\lambda|^{-N-0.5}) = e^{\lambda} + O(|\lambda|^{-N})  \eqno (4.5)
%\end{equation}
$$

\noindent
if $\lambda \rightarrow \infty,~~|arg \lambda| \le \frac{\pi}{2}$ 
for any integer $N$. Nevertheless we have to use the function $l(\lambda)$
in our calculations in Sect.~\ref{sect:central} and
Sect.~\ref{sect:confidence}. We consider it as 
a mathematical trick
to illustrate a possibility of constructing confidence intervals 
numerically for the real value $\hat x$.

Another approaches are also possible. 
At first, if we introduce a prior 
$\displaystyle g(\lambda) = \frac{e^{\lambda}}{\nu(\lambda)}$, 
then we have equality 
$\displaystyle \int_0^{\infty}{g(\lambda) f(x,\lambda)dx} = 1$
by natural way. 
Also we can numerically transform the 
function $f(x,\lambda)$ in the interval $x\in(0,1)$ so that 

$$
%\begin{equation}
\displaystyle \int_0^{\infty}{f(x,\lambda)dx} = 1,~~
E\xi = \displaystyle \int_0^{\infty}{xf(x,\lambda)dx} = \lambda,~~
D\xi = \displaystyle
\int_0^{\infty}{(x-E\xi)^2f(x,\lambda)dx} = \lambda  \eqno (4.6)
%\end{equation}
$$
\noindent
for any $\lambda$. In these cases we can construct the confidence interval
without introducing $l(\lambda)$.

Let us construct the central confidence interval for the continued
analogue of Poisson distribution using the function $l(\lambda)$.

\section{The Central Confidence Intervals for
the Continued Analogue of Poisson Distribution.}
\label{sect:central}

As we have noticed, for the discrete Poisson distribution
the mean coincides with the estimation of parameter value, 
i.e. $\hat \lambda = \hat x$.
This is not true for a small value of $\lambda$ in the considered case
(see Table~\ref{table:3}). That is why in order to find the estimation of
$\hat \lambda(\hat x)$ for a small value $\hat x$ it is necessary to
introduce the correction in accordance with Table~\ref{table:3}. 
Let us construct
the central confidence intervals using a conventional method assuming that

$$
%\begin{equation}
\displaystyle \int_{\hat x}^{\infty}{f(x,\hat \lambda_1)dx} = \frac{\alpha}{2} 
  \eqno (5.1)
%\end{equation}
$$

\noindent
for the lower bound $\hat \lambda_1$ and

$$
%\begin{equation}
\displaystyle
\int_{l(\hat \lambda_2)}^{\hat x}{f(x,\hat \lambda_2)dx} = \frac{\alpha}{2}
 \eqno (5.2)
%\end{equation}
$$

\noindent
for the upper bound $\hat \lambda_2$ of the confidence interval.

\begin{figure}
\epsfig{file=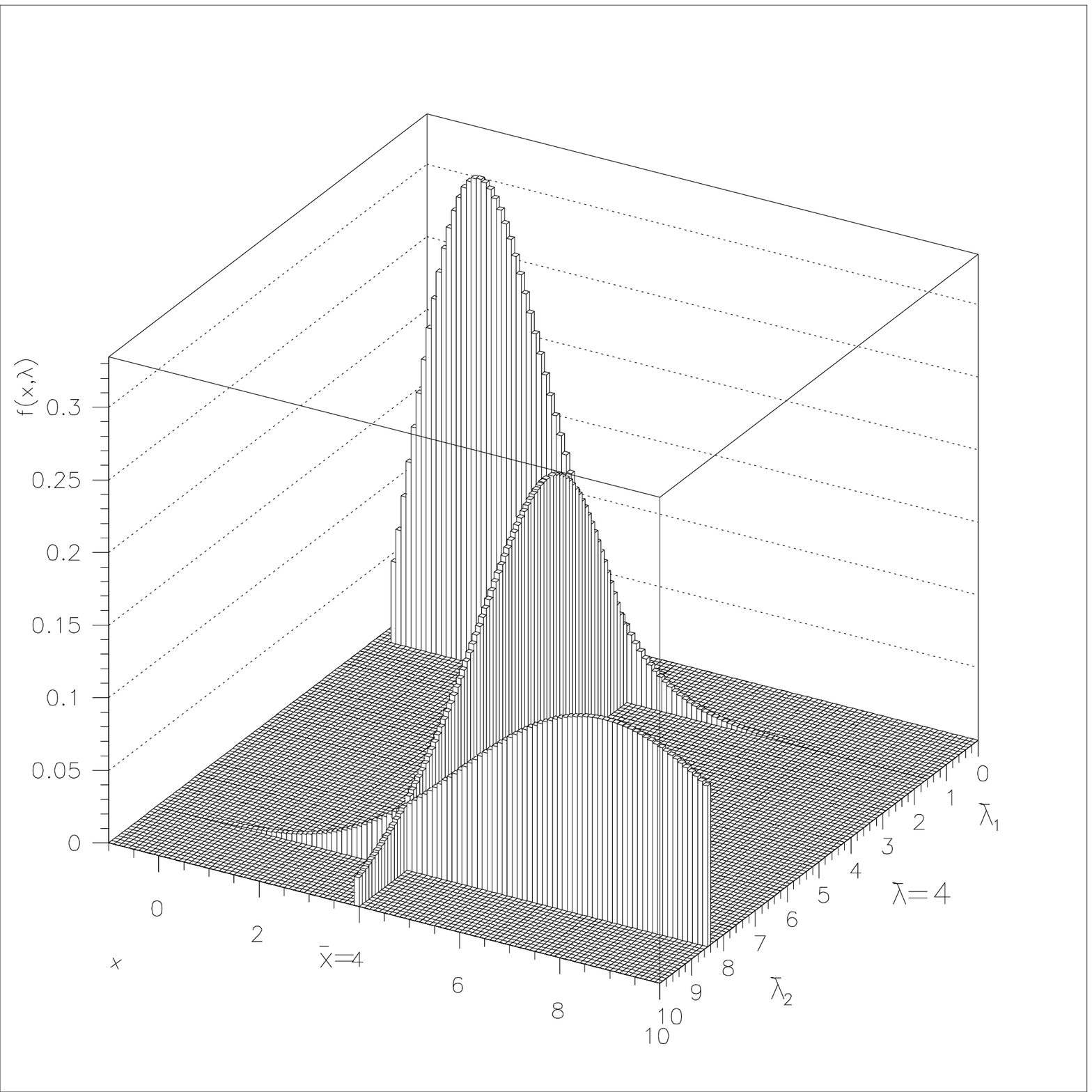,width=15cm}
%  \centering \includegraphics[height=0.4\hsize]{surf4.eps}
%\centerline{\psfig{figure=surf4.eps,height=6.2cm}}
\caption{\small The probability densities $f(x,\lambda)$ 
of continued analogue Poisson distribution for $\lambda$'s
determined by the
confidence limits $\hat \lambda_1$ and $\hat \lambda_2$
in case of the observed number of events $\hat x = 4$ and the
probability density of Gamma distribution with parameters $a=1$ and
$x+1=\hat x+1=5$.}
\label{fig:5}
\end{figure}

Fig.~\ref{fig:5} shows the introduced 
distributions (Sect.~\ref{sect:general}) with
parameters defined by the bounds of confidence interval 
$(\hat \lambda_1 = 1.638, \hat \lambda_2 = 8.493)$
for $\hat x = \hat \lambda = 4$ and the
Gamma distribution with parameters $a = 1$, $x+1 = \hat x+1 = 5$. 

%%%%%%%%%%%%%%%%%%%%%%%%%%fig.5

The bounds of confidence interval with a 90\% confidence level
for the parameter of continued analogue of Poisson distribution for different
observed values $\hat x$ (first column) were calculated and are given 
in the second column of Table~\ref{table:4}.

\setlength{\tabcolsep}{4pt}

\begin{table}
\begin{center}
\caption{90\% C.L. intervals for the Poisson signal mean $\lambda$
for total events observed $\hat x$.}
\label{table:4}
\begin{tabular}{|c|rl|rl|}
\hline
    & bounds   & (Section 5) & bounds & (Section 6)  \\ 
$\hat x$&$\hat \lambda_1$&$\hat \lambda_2$&$\hat \lambda_1$&$\hat \lambda_2$\\
\hline
 0.000 & 0.121E-08 &  2.052 &0.0       &  2.303  \\
 0.001 & 0.205E-08 &  2.054 &0.0       &  2.304   \\
 0.002 & 0.292E-08 &  2.056 &0.0       &  2.306   \\
 0.005 & 0.666E-08 &  2.061 &0.0       &  2.311    \\
 0.02 & 0.218E-06 &  2.098 & 0.0       &  2.337   \\
 0.05 & 0.765E-05 &  2.166 &  1.66E-05 &  2.389     \\
 0.10 & 0.137E-03 &  2.275 &  2.23E-05 &  2.474 \\
 0.20 & 0.186E-02 &  2.490 &  6.65E-05 &  2.642   \\
 0.30 & 0.696E-02 &  2.692 &  1.49E-04 &  2.806   \\
 0.40 & 0.161E-01 &  2.891 &  2.60E-03 &  2.969   \\
 0.50 & 0.295E-01 &  3.084 &  5.44E-03 &  3.129  \\
 0.60 & 0.466E-01 &  3.269 &  1.35E-02 &  3.290  \\
 0.70 & 0.673E-01 &  3.450 &  2.63E-02 &  3.452   \\
 0.80 & 0.911E-01 &  3.629 &  4.04E-02 &  3.611   \\   
 0.90 & 0.1179     &  3.804 & 6.12E-02 &  3.773   \\   
  1.0 & 0.1473     & 3.977  & 8.49E-02 &  3.933  \\
  1.5 & 0.3257     & 4.800  & 0.2391   &  4.718     \\
  2.0 & 0.5429     & 5.582  & 0.4410   &  5.479  \\
  2.5 & 0.7896     & 6.340  & 0.6760   &  6.220     \\
  3.0 & 1.056      & 7.076  & 0.9284   &  6.937  \\
  3.5 & 1.340      & 7.792  & 1.219    &  7.660     \\
  4.0 & 1.638      & 8.493  & 1.511    &  8.358  \\
  4.5 & 1.946      & 9.188  & 1.820    &  9.050    \\
  5.0 & 2.264      & 9.869  & 2.120    &  9.714   \\
  5.5 & 2.590      & 10.55  & 2.453    &  10.39    \\
  6.0 & 2.924      & 11.21  & 2.775    &  11.05  \\   
  6.5 & 3.264      & 11.87  & 3.126    &  11.72   \\
  7.0 & 3.609      & 12.53  & 3.473    &  12.38  \\
  7.5 & 3.961      & 13.18  & 3.808    &  13.01     \\
  8.0 & 4.316      & 13.82  & 4.160    &  13.65  \\
  8.5 & 4.677      & 14.46  & 4.532    &  14.30     \\
  9.0 & 5.041      & 15.10  & 4.905    &  14.95 \\
  9.5 & 5.406      & 15.73  & 5.252    &  15.56    \\   
  10. & 5.779      & 16.36  & 5.640    &  16.21  \\   
  20. & 13.65      & 28.49  & 13.50    &  28.33 \\
\hline
\end{tabular}
\end{center}
\end{table}

\setlength{\tabcolsep}{1.4pt}

As a result (Table~\ref{table:4}) the suggested approach 
allows one to construct confidence intervals for any real and integer 
values of the observed number of events for the values of
parameter $\lambda > 3$. Table~\ref{table:4} 
illustrates that the left bound of central 
confidence intervals is not equal to zero for small $\hat x$. It 
shows that in this case a central confidence interval is not suitable.

To anticipate a little, 
note that 90\% of the area of Gamma distributions with the parameter
$x+1 = \hat x+1$ are contained inside the constructed 90\% confidence intervals
for the observed value $\hat x$. However, for small values 
of $\hat x$ we have got 
values of the area close to 88\%, i.e. less than 90\%.  

The main goal of the 
proposed construction is to demonstrate a possibility of using 
a continuous two-dimensional function (4.1) for the construction
of confidence intervals in a frequentist meaning.

\section{Confidence Intervals for the Parameter of Poisson 
Distribution in case of the real value of observed number of events.}
\label{sect:confidence}

As follows from formulae $(2.7)$ and $(2.8)$ (see Figs.~\ref{fig:1}-\ref{fig:2}) 
the probability of true value of parameter of Poisson distribution to be 
$\lambda$ in case of observed integer value $\hat x \ge 0$ 
distributes in accordance with the Gamma distribution 
with the parameters $a = 1$ and $x + 1 = \hat x + 1$, i.e. 
according to formula $(2.4)$

\begin{center}
$P(\lambda|\hat x) = 
\displaystyle \frac{\lambda^{\hat x}}{\hat x!} e^{-\lambda}$. 
\end{center}

\noindent
The possibility of constructing the 
continued analogue of Poisson distribution
suggests 
to assume that the Gamma distribution of true value of the parameter
$\lambda$ takes place in case of 
the real value $\hat x \ge 0$ too (Figs.~\ref{fig:3}-\ref{fig:5}).
This supposition allows one 
to choose a confidence interval (for example) of a minimum length 
of all the possible confidence intervals of the given confidence level. 
The bounds of 
minimum length area, containing 90\% of the corresponding area of 
probability density of
Gamma distribution, were found numerically for several 
values of $\hat x$. 
We took into account that $x = \hat x$ and required that 
$\lambda_1 \ge 0$.
The results are presented in the third column of Table~\ref{table:4}.

\section{Conclusion}

In the paper the frequentist approach to construct the confidence 
interval for Poisson distribution parameter is considered. It is shown
that the formula $P(\lambda_1~\le~\lambda~\le~\lambda_2|\hat x) = 
P(n \le \hat x|\lambda_1) - P(n \le \hat x|\lambda_2)$ is a 
self-consistent definition of the confidence interval in this case. 
It means that the probability of true value of parameter of Poisson 
distribution to be equal to the value of $\lambda$ in the case of 
$m$ measurements $\hat x_1, \hat x_2, \dots, \hat x_m$ 
has probability density of Gamma distribution 
$\Gamma_{m,1+\sum_{i = 1}^{m}{\hat x_i}}$. 
%one
%measurement $\hat x$ has probability density of Gamma distribution 
%$\Gamma_{1,1+\hat x}$. 
The results of constructing the frequentist confidence intervals 
for the parameter $\lambda_s$ of Poisson distribution
for the signal in the presence of background with the known value
of parameter $\lambda_b$ are presented. It is shown that
the used procedure has both the Bayesian and frequentist interpretations.
Also the attempt of introducing a continued
analogue of Poisson distribution for the construction of 
confidence intervals for the parameter $\lambda$ of Poisson distribution
is discussed. Two approaches are considered. 
Confidence intervals for different integer and real 
values of the number of the observed events for the Poisson process 
in the experiment
with a given integral luminosity are constructed. 

We are grateful to V.A.~Matveev and V.F.~Obraztsov
for the interest to this work and for valuable comments. 
We wish to thank S.S.~Bityukov, A.V.~Dorokhov, V.A.~Litvine and 
V.N.~Susoikin for useful discussions. We would like
to thank E.A.~Medvedeva and E.N.~Gorina 
for the help in preparing the article.

This work has been supported by  
grant INTAS-CERN 99-377 and grant INTAS-CERN 00-440.

\end{document}